\documentclass[%
 floatfix,
 aip,
 amsmath,amssymb,
 reprint,%
]{revtex4-2}
\usepackage{chemformula} % Formula subscripts using \ch{}
\usepackage{graphicx}% Include figure files
\usepackage{dcolumn}% Align table columns on decimal point
\usepackage{bm}% bold math
\usepackage{physics}

\usepackage[utf8]{inputenc}
\usepackage[T1]{fontenc}
\usepackage{mathptmx}
\usepackage{verbatim}

\begin{document}

\preprint{AIP/123-QED}
%TC:ignore
\title{Substrate-controlled dynamics of spin qubits in low dimensional van-der-Waals materials}
% Force line breaks with \\

\author{Mykyta Onizhuk}
\affiliation{Department of Chemistry, University of Chicago, Chicago, IL 60637, USA}
\author{Giulia Galli}%
 \email{gagalli@uchicago.edu.}
 \affiliation{Department of Chemistry, University of Chicago, Chicago, IL 60637, USA}
 \affiliation{Pritzker School of Molecular Engineering, University of Chicago, Chicago, IL 60637, USA}
 \affiliation{Materials Science Division and Center for Molecular Engineering, Argonne National Laboratory, Lemont, IL 60439, USA}

\date{\today}% It is always \today, today,

\begin{abstract}
We report a theoretical study of the coherence dynamics of spin qubits in two-dimensional materials (2DMs) and van-der-Waals heterostructures, as a function of the host thickness and the composition of the surrounding environment. We focus on \ch{MoS_2} and \ch{WS_2}, two promising systems for quantum technology applications, and we consider the decoherence arising from the interaction of the spin qubit with nuclear spins. We show that the Hahn-echo coherence time is determined by a complex interplay between the source of decoherence in the qubit host and in the environment, which in turn determines whether the noise evolution is in a classical or quantum mechanical regime. We suggest that the composition and thickness of van-der-Waals heterostructures encapsulating a qubit host can be engineered to maximize coherence times. Finally, we discuss how quantum sensors may be able to probe the dynamics of the nuclear bath in 2DMs.
% The point I think we're making in the conclusions is that the quantum sensors can be used to probe the dynamics of nuclear bath in 2DM 
\end{abstract}

\maketitle
%TC:endignore

%TC:ignore
% \detailtexcount{main}
%TC:endignore

%%% Introduction

In the last few years, two-dimensional materials (2DMs) have attracted widespread attention in the field of quantum technologies\cite{Liu2019}, with potential applications as spin quantum dot qubits \cite{PhysRevB.86.085301, Volk2013} and single-photon emitters \cite{Chejanovsky2016, Koperski2015, MoS2_SE}. Recently, the coherent control of atomic defects in a 2DM has been reported for a monolayer of hexagonal boron nitride (h-BN) \cite{Exarhos2019, Gottscholl2020, gottscholl2020room}.
Moreover, theoretical studies have predicted a significant increase in the coherence time ($T_2$) of defect-based qubits in monolayers compared to their bulk counterparts \cite{Ye2019}.
% and pointed out that isotopic purification is a more efficient technique to increase $T_2$ in two than three dimensions .

However, the presence of the environment surrounding a 2D host may change its properties and hence those of the qubit; therefore environmental effects are expected to play an important role in the control and design of spin defects in two dimensions. For example, the nature of the substrate significantly alters the photoluminescence of \ch{WS_2} \cite{McCreary2016}, and the band gap of a \ch{MoS_2} monolayer, with variations of more than 1 eV (between 1.23-2.65 eV)\cite{Gyu2016}. In some cases, the presence of the environment may be beneficial for 2DM applications:
combining several layers of 2DMs leads to materials with interesting properties for nanoelectronics\cite{Fiori2014, Lee2020}, including atomic-scale transistors \cite{Britnell947} and memory units \cite{Bertolazzi2013}.

% for example, graphene may be turned into a semiconductor with a small gap of 0.5 eV when placed on a SiC substrate\cite{PhysRevLett.115.136802}, thus becoming a potential spin qubit host. Furthermore,

In this work, we present a theoretical investigation of the impact of the environment on the quantum dynamics of defect-based qubits in 2DMs and van-der-Waals heterostructures. We consider spin defects in wide-band-gap transition metal dichalcogenides \ch{WS_2} and \ch{MoS_2}, which are promising platforms for optoelectronic applications\cite{Choi2017} and quantum emitters \cite{Aharonovich2016}. We focus on a single source of decoherence, the interaction of the spin-defect with the surrounding nuclear spin bath, known to be the limiting factor for the coherence time of many solid-state qubits\cite{Yang_2016, Chirolli2008}.

%%% Model
Assuming a pure dephasing regime, we model the spin dynamics of the qubit using the cluster correlation expansion (CCE) method\cite{PhysRevB.78.085315, PhysRevB.79.115320}, which has been shown to yield accurate results for numerous systems \cite{Ma2014, PhysRevB.89.045403, Seo2016, PRXQuantum.2.010311}. We model Hahn-echo experiments, and the coherence time is obtained from the decay of the coherence function $L(t)$, defined as a normalized off-diagonal element of the density matrix $\hat \rho $ of the qubit:
\begin{equation}
    L(t) = \left|\frac{\bra{0}\hat \rho (t) \ket{1}}{\bra{0}\hat \rho (0) \ket{1}}\right|
\end{equation}
The structure of suitable spin defects in \ch{WS_2} and \ch{MoS_2} is still an open question\cite{Lin_2016}. Hence we simply consider a model defect with spin-1 and $\ket{0}$ and $\ket{- 1}$ as qubit levels, similar to those of optically addressable qubits in 3D materials, e.g. the \ch{NV^-} center in diamond \cite{Kennedy2003} or the divacancy in SiC \cite{Bourassa2020}. Furthermore, we assume that the electronic energy levels associated with the defect are localized within one unit cell, and the spin defect interacts with the nuclear bath as a magnetic point dipole. We compute the quadrupole tensors for the nuclei with spin $\ge 1$ using density functional theory with the PBE functional \cite{PhysRevLett.77.3865}, and the GIPAW module \cite{GIPAW} of the Quantum Espresso code \cite{Giannozzi2009}. We assume that for sparse baths the quadrupole interactions between nuclear spins and the electric field gradient are the same as those in the pristine material.

 \begin{figure}
    \centering
    \includegraphics[scale=1]{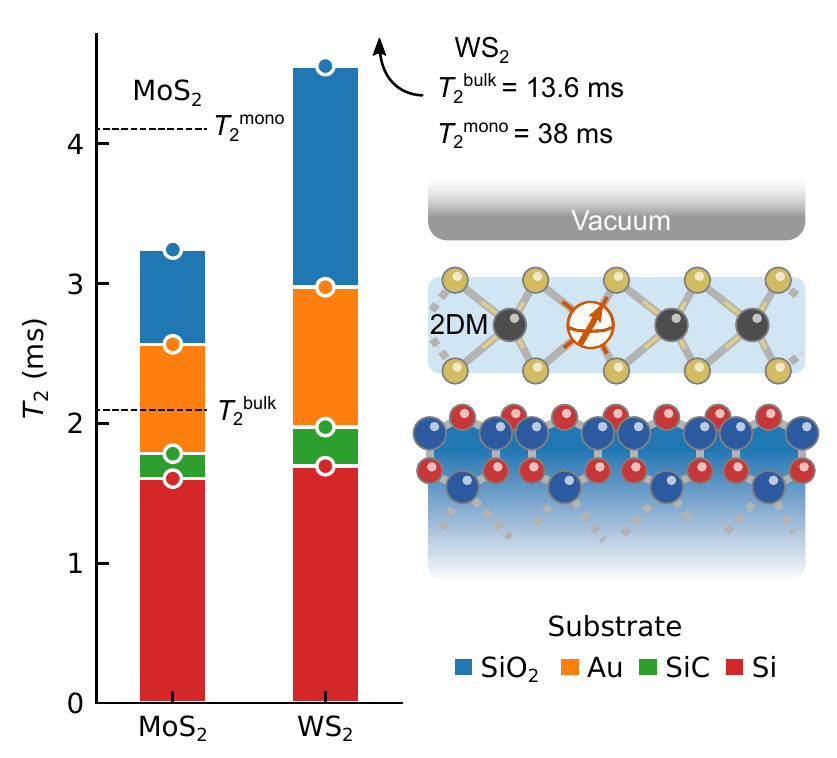}
    \caption{Left: Computed Hahn-echo coherence time ($T_2$) of 2DMs as a function of different substrates listed on the right.
    Distance between 2DM and the surface is $\sim$ 0.3 nm \cite{Velicky2018, PhysRevB.87.165402, Wang2019}. Applied magnetic field is perpendicular to the surface. Right: Representation of the model system.}
    \label{fig:substrates}
\end{figure}

\begin{figure*}
    \centering
    \includegraphics[scale=1]{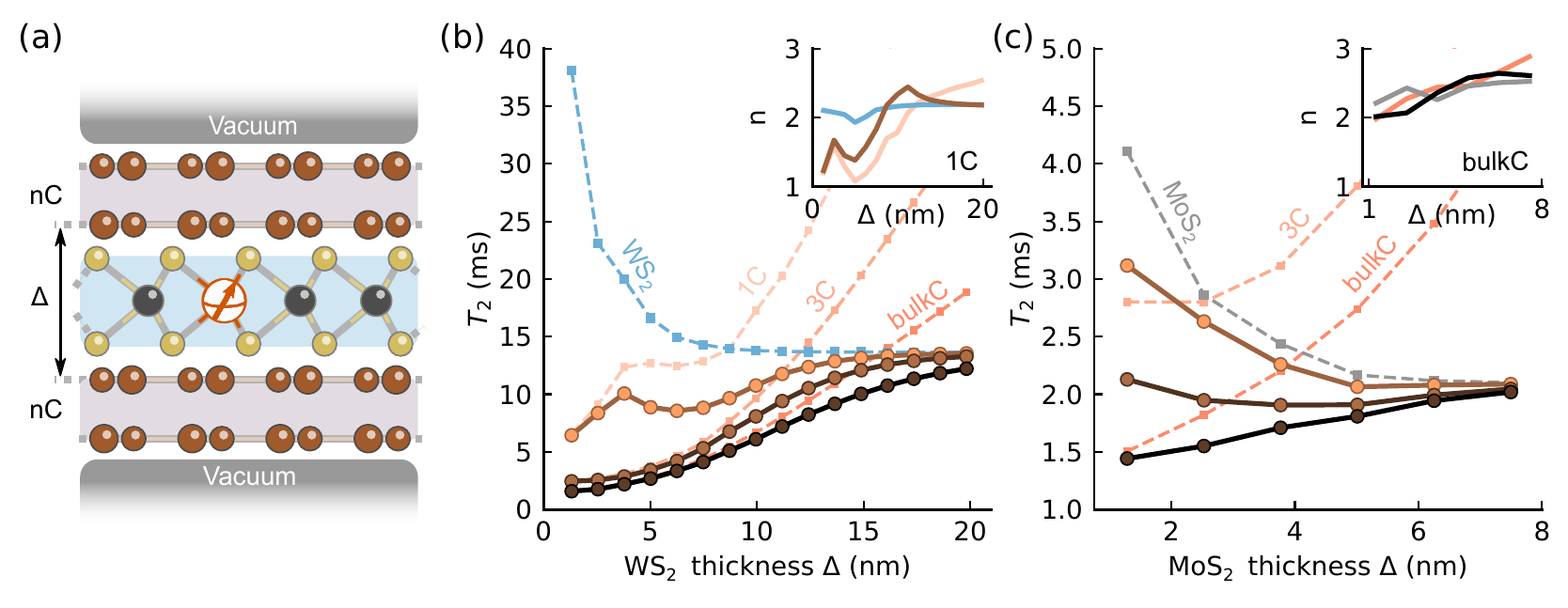}
    \caption{Computed Hahn-echo coherence time ($T_2$) of qubit located in low-dimensional heterostructure. (a) Schematic representation of the model system used in calculations. The qubit is located in the middle of the inner layer of thickness $\Delta$ (b), (c) $T_2$ of the C/\ch{WS_2}/C (C/\ch{MoS_2}/C) heterostructure depicted with solid lines. Dotted lines show the contributions of the host \ch{WS_2} (\ch{MoS_2}) and graphite substrate with different numbers of layers. The insets show the compressed exponential factor $n$ for host contribution in blue (grey), graphene (bulk graphite) substrate in orange, and the whole system in brown (black) as a function of $\Delta$. Graphite-host distances available from refs. \cite{Zhang2017, Wang2015_jpcc}.} 
    \label{fig:c_sand}
\end{figure*}

%%% Substrates
We start by discussing our results for the \ch{MoS_2} and \ch{WS_2} 2DMs.
In the absence of any interaction with the environment, Ye et al. \cite{Ye2019} predicted that $T_2$ of a spin qubit in \ch{MoS_2} increases by a factor of $\simeq$ 2 ($T_2^\text{mono}=2.2$, $T_2^\text{bulk}=1.2$ ms) when the nuclear bath of the host is reduced from 3D to 2D. The calculations of Ref.\cite{Ye2019} neglected the quadrupole term in the spin Hamiltonian. When including such term, we found $T_2$ increased further ($T_2^\text{mono} = 4.1$ ms), while the ratio between $T_2$ in 2D and 3D remains the same as previously reported. In natural \ch{WS_2}, the nuclear bath contains a smaller number of nuclear spins than in \ch{MoS_2}, with a relatively small gyromagnetic ratio. Hence the nuclear spin-induced decoherence in \ch{WS_2} is an order of magnitude slower than in \ch{MoS_2}, and the increase in coherence time with reduced dimensionality becomes more significant: $T_2^\text{mono}=38$ ms, 3 times higher than $T_2^\text{bulk}=13.6$ ms. 

Next, we investigate how coherence times vary when \ch{MoS_2} and \ch{WS_2} are deposited on a substrate. We consider Au (111), Si (111) and \ch{SiO_2} (ideal siloxane-terminated surface) \cite{Shi2014, Wood2020, Shi_2016, Hasani2019}. These substrates have been used in spectroscopic measurements \cite{Wood2020}, and applications of 2DMs in catalysis\cite{Shi2014, Shi_2016} and electronics \cite{Hasani2019}.
We neglect the reconstruction of the surface of the substrate and we assume that the quadrupole couplings are the same as those of the pristine material in vacuum.

Under an applied magnetic field, the contributions to coherence times of the different species of the nuclear bath are decoupled \cite{Seo2016}; hence:
\begin{equation}\label{eq:factorization}
    L=L^{S}L^{2DM}
\end{equation}
where $L^{S}$ and $L^{2DM}$ are the contributions of the substrate (S) and of the 2DM host, respectively.
We find that the nuclear baths of both the 2DM and the substrate may act as limiting factors to the qubit coherence. In particular, in \ch{WS_2} the qubit dynamics is completely determined by the substrate nuclear bath, and $T_2$ is significantly smaller than the one of the qubits in bulk \ch{WS_2} (Fig. \ref{fig:substrates}).
The coherence time depends on the nature of the substrate. The longest and shortest coherence times are obtained for \ch{SiO_2} and the Si (111) surface, respectively. These results can be understood by noticing that the Si substrate has the highest concentrations of \ch{^{29}Si} spins, compared to \ch{SiO_2} and \ch{SiC}; the latter exhibits however an additional source of decoherence given by the \ch{^{13}C} bath.
% [GG: can we comment on why we think this is the case?].
We also note that while gold contains the highest concentration of nuclear spins (100\% \ch{^{197}Au} with $s=\sfrac{3}{2}$), the small gyromagnetic ratio and large separation of the nuclei lead to a moderate influence of the substrate on the coherence time of the 2D material, with  $T_2 = 3$ ms.

In \ch{MoS_2}, on the other hand, both the substrate and host contribute significantly to the qubit decoherence. We find that an enhancement of T$_2$ from the reduced dimensionality persists only for the \ch{SiO_2} and gold substrates. In the presence of a Si (111) surface, the $T_2$ of a qubit in \ch{MoS_2} and \ch{WS_2} is almost identical.
% [GG: a simple explanation of why both bath and host contribute in the case of MoS2?]

%Therefore, we conclude that the choice of the host becomes relevant only when the decoherence rate is similar between the two sources.
% [GG: not sure I understand this: it was relevant also in the case of WS2...need to explain this better]
% This is rather a sub conclusion on substrate section. We can remove it altogether

We note that the presence of paramagnetic defects both on the surface of the substrate and in the 2DM itself may significantly impact decoherence rates and limit the value of $T_2$. To decrease the number of paramagnetic impurities in the 2D host, advanced experimental techniques are being developed \cite{Rhodes2019}. In general, to eliminate surface charges, it is desirable to reduce interfacial reconstructions as much as possible, and one way to achieve this objective is the use of van-der-Waals bonded materials as substrates. \cite{PhysRevLett.121.247701, Rhodes2019}.
% [GG: a short explanation of why; this statement appears a bit out of the blue]

%%% Sandwich systems.
Hence we turn to consider heterostructures of van-der-Waals bonded materials, which are emerging as promising platforms for 2D-based photonics. For example, a heterostructure of layered graphene and \ch{WS_2} was recently used to realize atomic defect-based photon emitters \cite{Schulereabb5988}.
% Next sentence can be significantly shortened if need be, but if limit allows I prefer to keep it.
To simulate the qubit dynamics in van-der-Waals bonded integrated systems, we focus on layered heterostructures with the spin qubit located in an inner layer, as shown in Fig. \ref{fig:c_sand}a.

We investigated the effect of the decoherence arising from the presence of a substrate (outer layer) and from the host (inner layer) of thickness $\Delta$; the coherence time is obtained by fitting the coherence function contribution (Eq. \ref{eq:factorization}) to the compressed exponential function, $L^M = \exp{\left[-(\sfrac{t}{T_2^M})^n\right ]}$, where $M$ denotes either the substrate (S) or the host 2DM.
We found that as $\Delta$ increases, the rate of the decoherence induced by the host bath increases (see Fig. \ref{fig:c_sand}(b) and Fig. \ref{fig:c_sand}(c) for \ch{WS_2} and \ch{MoS_2}, respectively). Instead the decoherence rate originating from the substrate nuclear bath decreases with the distance of the qubit from the substrate (see Fig. \ref{fig:c_sand}a and \ref{fig:c_sand}b, where $m$C denotes the number of Carbon layers). The combination of these two factors may result in a non-monotonic behavior of the total coherence time of the heterostructure as a function of $\Delta$.

Fig. \ref{fig:c_sand}(b) shows our results for C/\ch{WS_2}/C heterostructures. For $\Delta \le 3$ nm,
the effect of the substrate completely supersedes the effect of the host nuclear bath. With increasing $\Delta$, $T_2^{WS_2}$ decreases, but $T_2^C$ increases. We find that for $\Delta \ge 5$ nm, $T_2^C$ is proportional to $\Delta^\alpha$, where $\alpha$ depends on the thickness of the outer layer; in particular $\alpha = 1.67$ for bulk graphite and $2.5$ for graphene. The interplay between the host- and substrate-induced decoherence leads to the appearance of a local maximum in the coherence time. When the thicknesses of the host material exceeds $\sim 15$ nm, the decoherence is essentially limited by that of the \ch{WS_2} bath.

Our results show that depending on the number of graphite layers in the substrate, the coherence time of monolayer \ch{MoS_2}-based qubits can be either smaller or larger than $T_2$ in bulk \ch{MoS_2} (Fig \ref{fig:c_sand}(c)). The presence of two similar sources of decoherence arising from the host and the graphite/3C substrates leads to a minimum of the total coherence time around 4 nm.
% [GG: there is confusion here about the thickness we are talking about: the host and the substrate: we need to make sure we define either a symbol to refer to the two or we clarify better; I am often confused myself]

As the distance from the graphite substrate increases, the nature of the substrate-induced decoherence process changes. In particular, the graphene-induced decoherence as a function of $\Delta$ exhibits the most complex behavior: we observe a transition from a Gaussian ($n = 2$) to an exponential decay ($n = 1$) of the coherence function, near the local maximum in $T_2^C$ and in $T_2$ of the C/\ch{WS_2}/C heterostructure. For large $\Delta$, $n$ approaches 3 for both graphene and bulk graphite environments (insets of Fig. \ref{fig:c_sand}b and c).

\begin{figure*}
    \centering
    \includegraphics[scale=1]{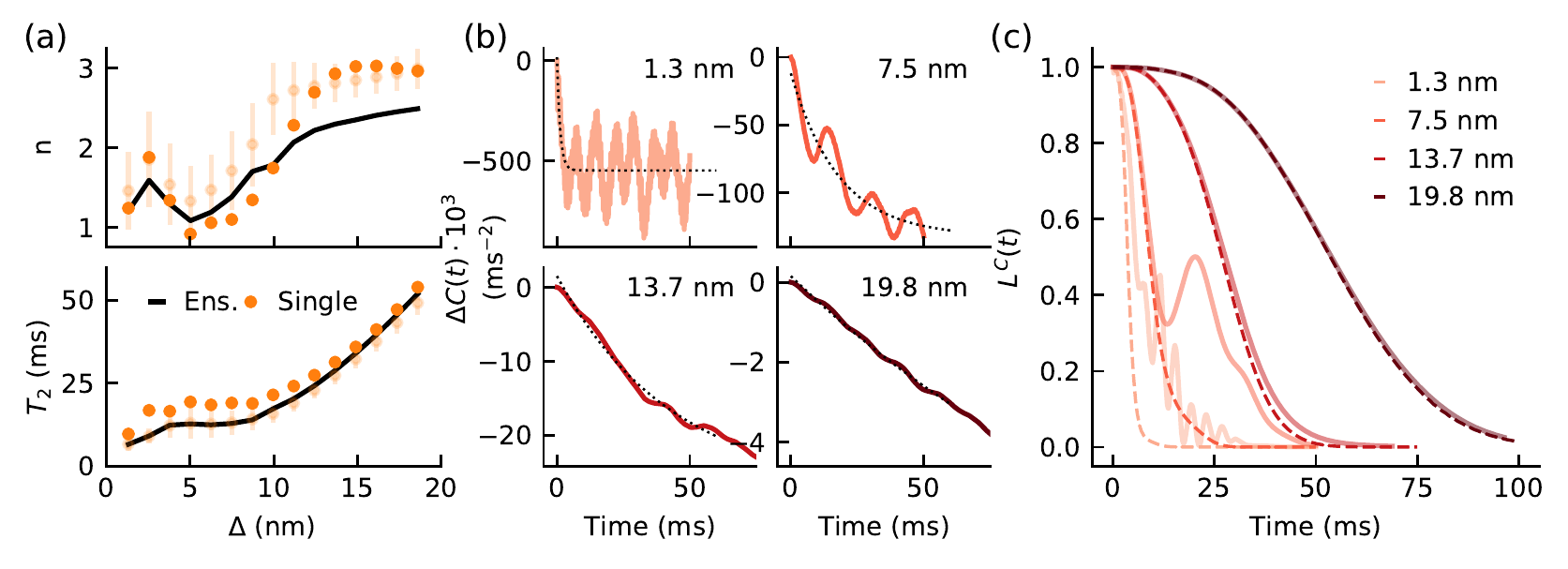}
    \caption{Graphene nuclear bath contribution to the qubit dynamics in the model heterostructures (Fig. \ref{fig:c_sand}). (a) Compressed exponential $n$ and coherence time $T^C_2$ as a function of the host thickness $\Delta$ for ensemble (black), mean value of single measurements $\pm1$ SD (light orange with error bars), and values for one specific configuration (orange). (b) Correlation function of the noise at different $\Delta$. Dotted lines show an exponential fit. (c) Graphene contribution $L^C$ at different $\Delta$ computed directly with the CCE method (solid lines) and reconstructed from correlation functions (dashed lines).}
    \label{fig:cnoise}
\end{figure*}

Fig. \ref{fig:cnoise} shows the decoherence induced by graphene monolayers in ensemble dynamics and for individual bath configurations. Our calculations show that the ensemble-averaged decay induced by the graphene nuclear bath has a smaller compressed exponent $n$ than the mean of the individual fitted decays (Fig. \ref{fig:cnoise}(a)), consistent with the predictions of stochastic noise models\cite{PhysRevB.102.134210, PhysRevB.77.245212}. However, we find that $n$ of the ensemble is not reduced by a factor of two, but only by $\sim 17\%$. Both ensemble and single spin dynamics show a non-monotonic variation of the coherence time with increased distance from the graphene substrate. We analyze below in detail the noise regimes of the Hahn-echo decay depending on the host thickness.

For $\Delta \ge 10$ nm, the dipolar coupling between nuclear spins in graphene is larger than the coupling of the graphene layer to the central spin.
In this case, it is reasonable to assume that the Hahn-echo decay is governed by a classical Gaussian noise, and hence by the Hamiltonian:
\begin{equation}
    \hat H^{\text{classical}}(t) = y(t)\eta(t)\hat S_z
\end{equation}
where $\hat S_z$ is the spin operator of the qubit electron spin, and $y(t) = \pm 1$ changes its sign each time a $\pi$ pulse is applied \cite{PhysRevB.90.115431}. $\eta(t)$ is a stochastic variable, corresponding to the magnetic noise. Then the coherence function can be computed as \cite{PhysRevA.86.012314}:

\begin{equation}\label{eq:corr}
    L(t) = \exp \left[\int_0^t{ C(u)F_t(u)} du \right]
\end{equation}

Where $C(t)=\langle \hat \eta_{\text{eff}}(t) \hat \eta_{\text{eff}}(0)\rangle$ is the correlation of the Overhauser field of the nuclear bath. $F_t(u)$ is a correlation filter function, defined as \cite{PhysRevA.86.012314}:
\begin{equation}
    F_t(u)=\int_u^{2t-u}y\left(\frac{v-u}{2}\right)y\left(\frac{v+u}{2}\right)dv
\end{equation}

We compute the noise correlation function with the CCE method \cite{PhysRevB.90.115431, PhysRevB.92.161403} from the autocorrelation function of the Overhauser field operator $\hat \eta_{\text{eff}}(t) = \sum_i {A_{||}\hat I_z^i (t)}$ where $A_{||}$ is hyperfine coupling. The correlation function for one random nuclear spin configuration is shown in Fig. \ref{fig:cnoise}(b) for different separations of graphene layers from the qubit.  The dotted black lines show the exponential fit $\Delta C(t) = b^2 ( e^{-t/\tau_C} - 1)$, where $\tau_C$ is the correlation time of the bath. We observe that $\tau_C$ increases with thickness from 1.2 ms to 150 ms for the largest separation considered here. The long correlation time at large $\Delta$ and the coherence decay of $L^C\approx \exp[-(\sfrac{t}{T_2})^3]$ agree well with stochastic model predictions for the slow evolution of the bath\cite{PhysRev.125.912, PhysRevB.67.033301}.

We further observe the emergence of a classical regime by reconstructing the coherence function from the noise correlation using Eq. \ref{eq:corr} (Fig. \ref{fig:cnoise}(c)). We find that the semiclassical approach faithfully reproduces the complete quantum mechanical evolution of the bath at separations between graphene layers larger than 10 nm. 

\begin{figure}[b]
    \centering
    \includegraphics[scale=1]{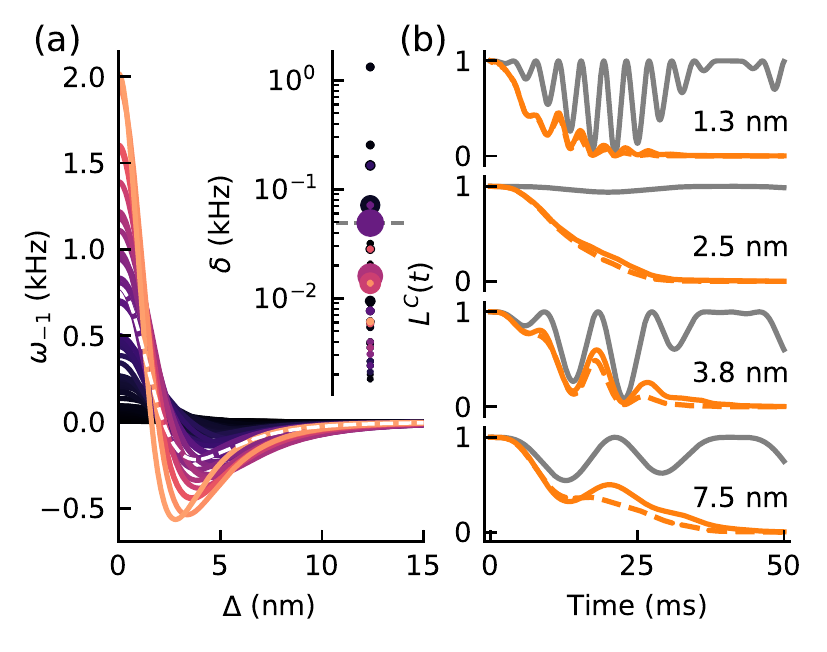}
    \caption{Dynamics of the pseudospins in a graphene nuclear bath. (a) Pseudospin frequency $\omega_{-1}$ for 100 spin pairs contributing to $L^C (t)$ as a function of the host thickness $\Delta$. Inset shows the corresponding flip rate $\delta$ for the given spin pairs. The symbol represents the maximum value of $\omega_{-1}$, and its size represents relative contribution. The dashed line corresponds to the pseudospin shown in (b). (b) Graphene contribution $L^C$ to the decoherence as a function of $\Delta$ computed with the CCE method (orange solid lines) and reconstructed from the pseudospin approximation (dashed lines). The grey solid lines show a single pseudospin contribution.}
    \label{fig:pseudo}
\end{figure}

For $\Delta \leq$ 10 nm, however, the hyperfine coupling is on a par or significantly larger than the average dipolar coupling between nuclear spins, and a classical treatment of the noise is no longer appropriate\cite{Yang_2016}. In this case, a complex decay of the coherence function is observed (Fig. \ref{fig:cnoise}c), which is not captured by Eq. \ref{eq:corr}.

An insight into this complex dynamics can be gained by analyzing the evolution of a single pair of spins $i$ and $j$. At sufficiently strong magnetic fields, only pairwise spin flip-flops are allowed ($\ket{\uparrow\downarrow} \leftrightarrow \ket{\downarrow\uparrow}$), and their dynamics can be mapped on a 2-level "pseudospin", governed by the Hamiltonian:
\begin{equation}\label{pseudo}
    \hat H^{\text{pseudo}} = \frac{\omega (t)}{2} \hat \sigma_z + \frac{\delta}{2} \hat{\sigma_x}
\end{equation}
where the frequency of the pseudospin, $\omega (t) = \omega_{-1} = \Delta A_{||}$,
is given by the difference in hyperfine coupling, if the central spin is in the $\ket{-1}$ state, or $\omega (t) = \omega_{0} = 0$ if the central spin is in the $\ket{0}$ state. $\sigma_i$ are Pauli matrices for spin-\sfrac{1}{2}. $\delta = \frac{\hbar \gamma_i \gamma_j}{2r_{ij}^3}(3\cos^2\theta-1)$ is obtained from the dipolar coupling of $i,j$ nuclear spins, where $\theta$ is the angle between the vector $r_{ij}$ connecting two nuclear spins and the external magnetic field.

The total coherence function can be obtained analytically \cite{PhysRevB.74.195301, Seo2016} as a product of the contributions of all spin pairs ${ij}$:

\begin{equation}\label{ps_L}
    L(t) = \prod_{ij}{\left[1 - \kappa \cdot
    \sin^2(\sqrt{\omega_{-1}^2+\delta^2}\frac{t}{4})
    \sin^2(\delta \frac{t}{4})\right]}
\end{equation}
where $\kappa = \frac{\omega_{-1}^2}{\omega_{-1}^2+\delta^2}$. Due to the reduced dimensionality of graphene and the sparse concentration of nuclear spins, only a small number of spin pairs exists in close proximity of the qubit and thus contributes to determining the coherence decay. (Fig. \ref{fig:pseudo}).

The dependence of $\omega_{-1}$ on the distance from the graphene layer can be computed from the hyperfine couplings of nuclear spins and is quite complex (Fig. \ref{fig:pseudo}(a)). As $\Delta$ increases, the frequency of pseudospins changes its sign and exhibits a local extremum. At $\omega_{-1} = 0$, the hyperfine couplings of two nuclear spins are the same, and the contribution of the spin pair to the coherence function becomes negligible (Fig. \ref{fig:pseudo}(b)). When the amplitude $\omega_{-1}$ reaches a local maximum, the spin pair's dynamics again significantly impacts the coherence function.

This unique behavior of the hyperfine couplings determines the non-monotonic dependence of $T_2^C$ and of the compressed exponent $n$ on the distance between the graphene layers and the qubit. We note the good agreement between the pseudospin prediction and the exact solution at small distances. For large values of $\Delta$, the pseudospin model gives an underestimation of the coherence time, due to the longer correlation time $\tau_C$ and to higher-order effects playing a dominant role.
% \begin{figure}
%     \centering
%     \includegraphics[scale=1]{figures/bn_full.pdf}
%     \caption{Hahn-echo coherence time of qubit located in BN/\ch{WS_2}/BN heterostructures. Dotted lines show contributions of decoupled nuclear baths. BN nuclear bath contribution is approximated at CCE2 level.}
%     \label{fig:ws2bn}
% \end{figure}

Finally, we note that there are cases in which the choice of the substrate in van-der-Waals structures can completely suppress the effect of the host at any host thickness of interest. As an example, we consider h-BN as a substrate, whose nuclear spin bath contains 100\% concentration of spins with a high gyromagnetic ratio. Using CCE calculations up to the second order, without quadrupole tensors included, we estimated the thickness of the qubit host at which the coherence time is determined purely by the host spin bath. We found a lower bound  of $\approx 20$ nm for \ch{MoS_2} and $\approx 80$ nm for \ch{WS_2}, respectively. 
% [GG: small final comment].

%%% Conclusions
In sum, we presented a theoretical study of the influence of the environment on the coherence time of spin qubits in 2D materials. We found that both the nuclear spins of the substrate and those of the host 2DM can act as sources of decoherence for the qubit. Our results show that the composition of van-der-Waals heterostructures encapsulating the qubit may be engineered to obtain longer coherence times.

Our calculations also revealed a complex behavior of the Hahn-echo coherence time as a function of the thickness of the 2D material hosting the qubit. For thin hosts ($\Delta \le 5$ nm) it is possible to identify specifically which  pairs of spins give rise to the oscillations observed in the Hahn-echo decay time. This result points at the   possibility of using the electron spin of the qubit as a sensor of  dipolar couplings within the 2D nuclear bath\cite{Abobeih2019}. For thicker hosts ($\Delta \ge 10$ nm), we observed a transition from a quantum to a classical regime of the induced decoherence, which might be revealed experimentally by using. e.g. a NV center in diamond as a quantum sensor\cite{Lovchinsky503, PhysRevX.10.011003}.

Finally we note that for the 2DMs investigated here, other decoherence channels (spin-orbit, spin-phonon interactions \cite{Xu2020, Wang2015, Yang2015}) exist, which may play a significant role at nonzero temperatures. Their contribution to decoherence needs to be carefully assessed in the future. And although we focused only on the interaction of the spin defect with the nuclear spin bath, our results point at the importance of the substrate and its composition in the design of  2DMs and van-der-Waals heterostructures for quantum applications.

%TC:ignore
\section*{Acknowledgements}
This work made use of resources provided by the University of Chicago’s Research Computing Center. This work was supported as part of the Center for Novel Pathways to Quantum Coherence in Materials, an Energy Frontier Research Center funded by the U.S. Department of Energy, Office of Science, Basic Energy Sciences. We thank Meng Ye, Yuxin Wang, and Siddhartha Sohoni for useful discussions.
\section*{Data Availability}
The data that support the findings of this study are available from the corresponding author upon reasonable request.
%TC:endignore
\bibliography{references}% Produces the bibliography via BibTeX.

\end{document}